%% file: paper.tex
\documentclass[%
 reprint,
 amsmath,amssymb,
 aps,
pra,
nofootinbib,
noeprint
]{revtex4-2}

\usepackage{placeins}
\usepackage{tabularx}
\usepackage{comment}
\usepackage{graphicx}
\usepackage{dcolumn}
\usepackage{bm}
\usepackage{color}
\usepackage{xfrac}
\usepackage{sidecap}

\newcommand{\ket}{\rangle}
\newcommand{\bra}{\langle}

\newcommand{\sixJ}[6] {\left\{\begin{array}{ccc} #1 & #2 & #3 \\ #4 & #5 & #6 \end{array}\right \}}
\newcommand{\threeJ}[6]{\left(\begin{array}{ccc} #1 & #3 & #5 \\ #2 & #4 & #6 \end{array}\right )}
\newcommand{\nineJ}[9]{\left\{\begin{array}{ccc} #1 & #2 & #3 \\ #4 & #5 & #6 \\#7 & #8 & #9 \end{array}\right \}}

\usepackage{rotating}

\usepackage[dvipsnames]{xcolor}

\usepackage{threeparttable}

\usepackage{here}

\begin{document}

\title{Engineering field-insensitive molecular clock transitions for\\ symmetry violation searches}

\author{Yuiki Takahashi}
\email{yuiki@caltech.edu}
\author{Chi Zhang}
\author{Arian Jadbabaie}
\author{Nicholas R. Hutzler}

 \affiliation{Division of Physics, Mathematics, and Astronomy, California Institute of Technology, Pasadena, California 91125, USA}

\date{\today}

\begin{abstract}

Molecules are a powerful platform to probe fundamental symmetry violations beyond the Standard Model, as they offer both large amplification factors and robustness against systematic errors.  As experimental sensitivities improve, it is important to develop new methods to suppress sensitivity to external electromagnetic fields, as limits on the ability to control these fields are a major experimental concern.  Here we show that sensitivity to both external magnetic and electric fields can be  simultaneously suppressed using engineered radio frequency, microwave, or two-photon transitions that maintain large amplification of CP-violating effects. By performing a clock measurement on these transitions, CP-violating observables including the electron electric dipole moment, nuclear Schiff moment, and magnetic quadrupole moment can be measured with suppression of external field sensitivity of $\gtrsim$100 generically, and even more in many cases.  Furthermore, the method is compatible with traditional Ramsey measurements, offers internal co-magnetometry, and is useful for systems with large angular momentum commonly present in molecular searches for nuclear CP-violation.

\end{abstract}

\maketitle

Precision measurements of heavy atomic and molecular systems have proven to be a powerful probe of physics beyond the Standard Model (BSM)~\cite{Safronova2018}. For example, searches for the electron’s electric dipole moment (eEDM) in ThO and HfF$^+$ probe charge-parity (CP) violating new physics at TeV energy scales~\cite{ACME2018, Cairncross2017, NewJILAResult}. Their high sensitivities rely on the enhancement of CP-violating observables in the internal molecular electromagnetic environment, combined with the advantageous experimental features arising from their unique molecular structures.  The sensitivity of these molecular CP-violation (CPV) searches will continue to improve as new methods such as laser cooling are developed to increase count rates and coherence times~\cite{Safronova2018,Hutzler2020Review,Alarcon2022Snowmass}.

The two most sensitive eEDM experiments~\cite{ACME2018, Cairncross2017, NewJILAResult} rely on a particular molecular structure ($^3\Delta_1$ state) which offers two critical features: a small magnetic moment, and ``internal co-magnetometers.''  The small magnetic moment of these states makes them largely insensitive to uncontrolled magnetic fields, which are a major challenge for EDM experiments with atoms~\cite{Regan2002} and neutrons~\cite{abel_measurement_2020}. Internal co-magnetometers give the ability to reverse the desired CPV energy shifts without changing lab fields, resulting in robustness against systematic effects arising from the applied electric and magnetic fields.

However, molecules which are sensitive to CPV and have this molecular structure are not laser coolable -- a feature which would be advantageous for improving sensitivity through advanced quantum control.  Polyatomic molecules offer internal co-magnetometers generically, including molecules which can be laser-cooled~\cite{Kozyryev2017PolyEDM,Hutzler2020Review}.  Laser-coolable, eEDM sensitive molecules have a single, metal-centered $s$-type valence electron~\cite{Isaev2016Poly,Fitch2021Review}, which means that they have magnetic moments on the order of the Bohr magneton $\mu_B$; this is $\gtrsim 100$ times larger than $^3\Delta_1$ states, and correspondingly more sensitive to magnetic fields.  A recent demonstration~\cite{Anderegg2023CaOHSpin} showed that it is possible to tune the magnetic sensitivity in polyatomic molecules to very low values while still maintaining eEDM sensitivity.  Another proposal to reduce field sensitivity is to use magnetically-insensitive ``clock'' transitions, which occur in many molecules~\cite{verma_electron_2020}.

In this manuscript, we discuss a generic method to engineer ``clock'' transitions which have reduced sensitivity to both magnetic \textit{and} electric fields while maintaining large sensitivity to CPV in molecules.  The basic idea behind these CPV-sensitive field-insensitive transitions (CP-FITs) is similar to magic conditions in atomic clocks~\cite{Derevianko2011_katori_lattice_clock, Ludlow2015} and precision spectroscopy in molecules~\cite{Prehn2021}. We find CP-FITs in a wide range of polyatomic and diatomic molecules with applications to search for the eEDM, nuclear Schiff moments (NSM), and nuclear magnetic quadrupole moments (MQM).  This technique is particularly useful for molecules with large nuclear spins and complicated hyperfine structure for nuclear CPV searches.  Furthermore, because CP-FITs do not involve traditional $M=0\rightarrow 0$ clock transitions, where $M$ is the angular momentum projection quantum number, driving the same transitions with the opposite signs of $M$ provides a co-magnetometry method to reject systematic errors.  Finally, since the states involved have static energy shifts sensitive to CPV, they can be probed with traditional Ramsey spectroscopy using RF, microwaves, or lasers.

The Hamiltonian governing the CPV energy shifts~\cite{Flambaum2014}
 in a molecule interacting with external fields is
\newcommand{\E}{\mathcal{E}}
\newcommand{\B}{\mathcal{B}}

\begin{multline}
H = -D\bm{n}\cdot\bm{\E}-g\mu_B\bm{S}\cdot\bm{\B} \\
+ W_dd_e \bm{S\cdot n} + W_Q\frac{Q}{I} \bm{I\cdot n} - W_M\frac{M}{2I(2I-1)} \bm{S\hat{T}n}.
\end{multline}
The first line is the external field interaction; $D$ is the molecular dipole moment, $\bm{n}$ is a unit vector along the molecular axis, $g$ is the magnetic $g$-factor, and $\bm{S}$ is the effective electron spin.  Here we define the molecular axis as the symmetry axis, along which the dipole moment lies. The second line is the CPV effects; $d_e$ is the eEDM,  $Q$ is the NSM, $\bm{I}$ is the nuclear spin, $M$ is the MQM, and $\bm{T}$ is the rank 2 tensor operator $T_{ij} = I_iI_j + I_jI_i  -  \frac{2}{3}\delta_{ij}I(I+1)$.
$W_d, W_Q$, and $W_M$ are sensitivities to eEDM, NSM, and MQM, which are determined by electronic structure, and are the same within the same electronic state to good approximation. On the other hand, $P_d \equiv \bra \bm{S\cdot n}\ket$, $P_Q \equiv  \frac{1}{I}\bra \bm{I\cdot n}\ket$, and $P_M \equiv \frac{1}{2I(2I-1)}\bra \bm{S\hat{T}n}\ket$ are measures of the eEDM, NSM, and MQM sensitivities which depend on each particular quantum state, and in general have different orientations of $\bm{S},\bm{I}$, and $\bm{n}$. In this manuscript, we refer to these state-dependent sensitivities $P_d, P_Q$, and $P_M$ as ``CPV sensitivities'' as we shall restrict our discussion to transitions within a given electronic state.

\begin{figure}[t]
    \includegraphics[width=0.48\textwidth]{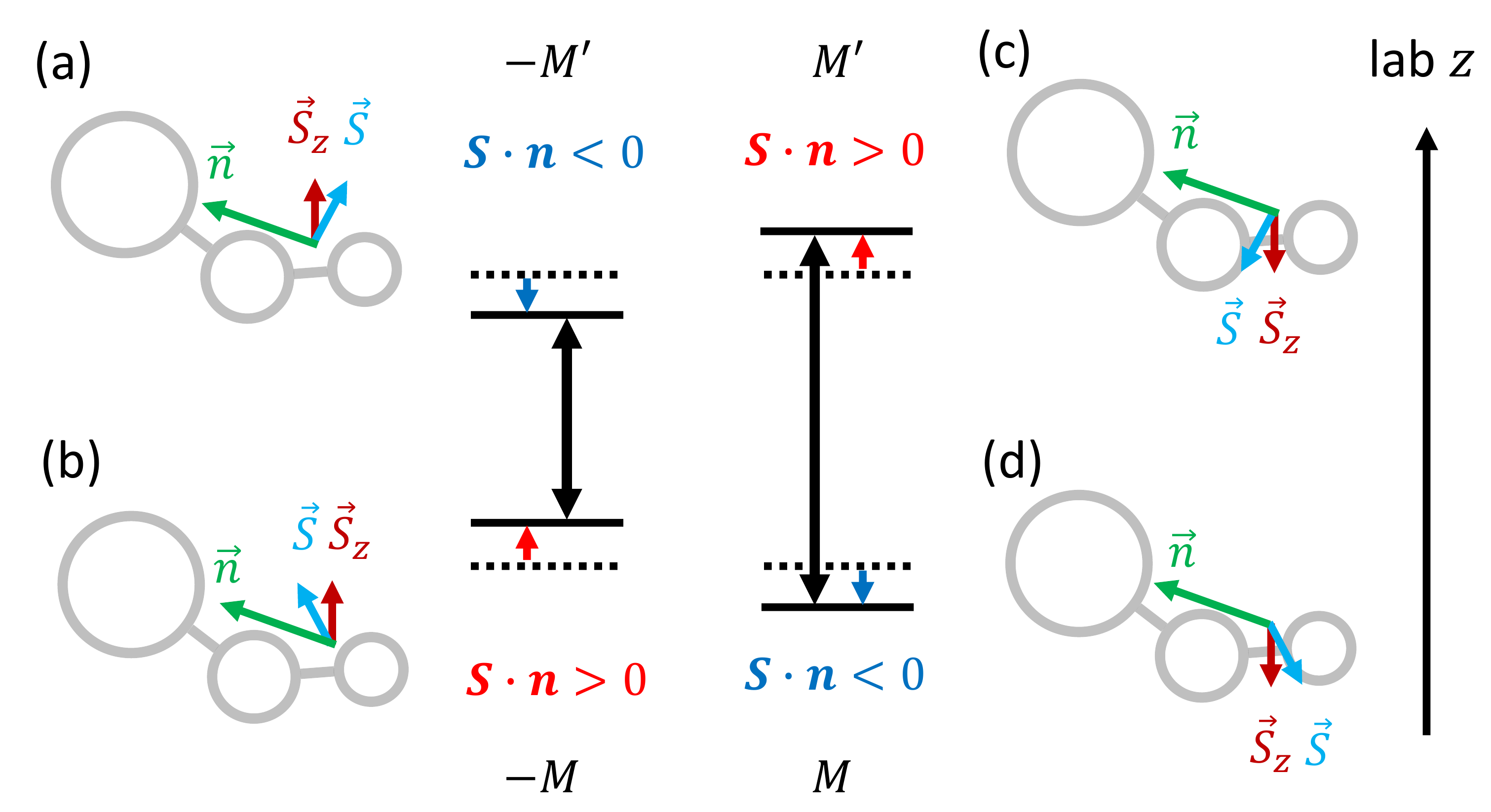}

\caption{\label{fig:intuitive} The CP-FITs with eEDM sensitivity in partially polarized molecules where the molecule axis $\bm{n}$ and the lab $z$ axis are not fully aligned. States (a) and (b) have the same orientation of $\bm{n}$ with respect to $z$ as well as the same projection $S_z$ of $\bm{S}$ on $z$, and thus have the same response to electric and magnetic fields. On the other hand, they have different orientations of $S$ on $\bm{n}$, and thus different eEDM energy shifts (short arrows). This results in transitions (long arrows) sensitive to eEDM interactions but highly insensitive to electric and magnetic fields.  The transition connecting states (c) and (d) is similar though it has the opposite eEDM shift, serving as an internal co-magnetometer. The same mechanism works for CP-FITs with NSM and MQM sensitivity.}
\vspace{-5mm}
\end{figure}

The existence of CP-FITs -- pairs of states having similar electric and magnetic field sensitivities but different CPV sensitivities -- can be intuitively understood as arising from the ability to access different orientations of the molecular $\bm{n}$ and lab $z$ axes. Figure \ref{fig:intuitive} shows a time-reversed pair of two CP-FITs in a partially polarized molecule where $\bm{n}$ and the lab $z$ axis are not fully aligned. The field sensitivities arise from the projections of $\bm{n}$ and $\bm{S}$ on the \textit{lab} axis whereas the CPV sensitivities arise from the projections of $\bm{S}$ and $\bm{I}$ on the \textit{molecular} axis. The interactions between $\bm{S}$, $\bm{I}$, and the angular momentum associated with the molecular axis (e.g, spin-rotation interaction), all of which generically exist in species with non-zero electron and nuclear spin, offer different orientations of $\bm{S}$ and $\bm{I}$ with respect to $\bm{n}$. Thus, when molecules are partially polarized, states with the same projections of $\bm{n}\cdot z$ and $\bm{S}\cdot z$ (field sensitivities) can have different projections of $\bm{S}\cdot\bm{n}$, $\bm{I}\cdot\bm{n}$, and $\bm{S\hat{T}n}$ (CPV sensitives). 

As the molecular Stark shifts begin to approach the magnitude of other splittings in the molecule, the coupling of the spins and dipole moment can transition from being molecule-quantized to lab-quantized, resulting in changes to the behavior of Stark and Zeeman shifts, including the appearance of pairs of states with the same absolute or differential (``magic'') field sensitivity. This has been observed in molecules~\cite{Prehn2021,Jadbabaie2023YbOH010}, and is, for example, the intuitive reason behind magnetically-insensitive states and transitions~\cite{Li2020} in alkali atoms.  In this manuscript, we show that these exist rather generically, and that we can find transitions that have large CPV-sensitivity.  

Although this mechanism is similar to what gives rise to eEDM-sensitive, magnetically-insensitive states in polyatomic molecules~\cite{Kozyryev2017PolyEDM,Anderegg2023CaOHSpin}, here we are exploring \textit{transitions} between states which may have large electromagnetic field sensitivity, but whose shifts are similar, making transition frequencies largely insensitive to electromagnetic fields while still sensitive to CPV.  Similar types of field-insensitive transitions in atoms and molecules have been explored in other contexts, including for clocks~\cite{Arnold2016}, quantum computing~\cite{Najafian_magic_for_qubitin_N2_2020}, searches for ultralight dark matter~\cite{Kozyryev2021}, and precision spectroscopy of polyatomic molecules~\cite{Prehn2021}, and in the latter case have proven their experimental power.

\begin{figure}[t]
    \includegraphics[width=0.48\textwidth]{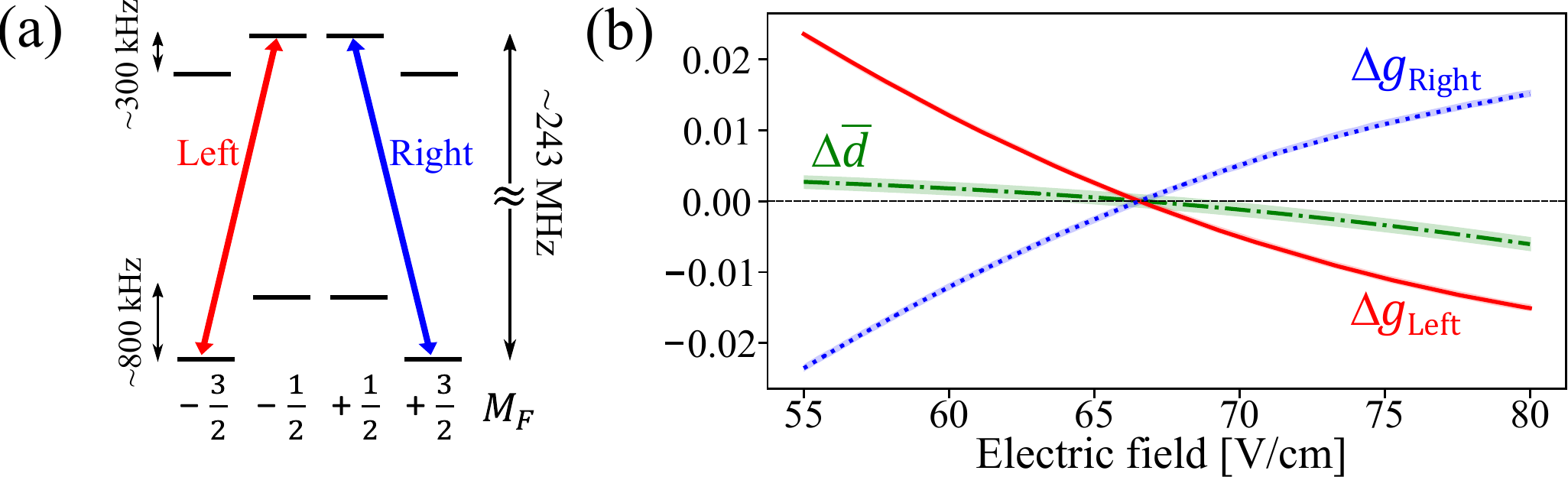}
\caption{\label{fig:magic173result} An example CP-FIT where both $\Delta \bar{d}$ and $\Delta g$ cross zero near 66.7 V/cm. (a): Level diagram of states and transitions involved. (b): $\Delta \bar{d}$ and $\Delta g$ as a function of the electric field. The shaded regions indicate the uncertainties from the measurement uncertainties of the spectroscopic parameters.} 
\end{figure}

A time-reversed pair of CP-FITs (that is, transitions between $M\leftrightarrow M'$ and $-M\leftrightarrow -M'$, see Figs.~\ref{fig:intuitive} and \ref{fig:magic173result}a) are degenerate in the absence of a magnetic field (and CPV), and have exactly the same projection $\bm{n}\cdot z$, but opposite projections of $\bm{S}$ and $\bm{I}$ on $\bm{n}$ -- that is, opposite CPV sensitivity. By comparing these two transition frequencies without changing lab fields, shifts due to the electric field are canceled while CPV energy shifts are not. Therefore, this scheme provides an opportunity for internal co-magnetometry, even for species and states without parity doublet structures.

As a specific example, we examine CP-FITs in the $^{173}$YbOH metastable bending mode of the electronic ground state, $\tilde{X}^2\Sigma_{1/2}^+(010)$. The combination of parity doublets from the mechanical bending vibration and the heavy, quadrupole-deformed $^{173}$Yb$(I=5/2)$ nucleus offers good sensitivity to the eEDM, MQM, and NSM~\cite{Kozyryev2017PolyEDM, denis_enhanced_2020, Maison2019}.  We shall find CP-FITs in this molecule, as well as a number of others.

We diagonalize the molecular effective Hamiltonian~\cite{Pilgram2021YbOHOdd, Jadbabaie2023YbOH010} including Stark and Zeeman terms to compute quantum state energy levels and magnetic, electric, eEDM, NSM, and MQM sensitivities at each electric field, as described in detail in the Supplemental Material.~\nocite{Hutzler2014Thesis, Steimle2019, Sawaoka_ZS_YbOH_2023, nakhate_pure_2019_no_doi} The nuclear spin and rotational Zeeman terms are neglected since their contributions are expected to be smaller by around three orders of magnitude than the electron spin term~\cite{Campbell_rotational_zeeman_1983, hirota_high-resolution_1985}. To represent the magnetic and electric sensitivity, the $g$-factor and dipole moment are computed by taking the first derivative of energies as a function of magnetic and electric fields, respectively.

We are able to find many values of the electric field supporting CP-FITs -- pairs of states which have large CPV sensitivity, small $g$-factor and dipole moment, and which differ by $\Delta M\leq 2$, which makes them easier to drive in the laboratory.  Table~\ref{tab:ybohmagic} lists several examples in $^{173}$YbOH, and Figure~\ref{fig:magic173result} shows one in detail.  We characterize these transitions in this and other molecules by computing the differential $g$-factor between the two states, $\Delta g=g_1-g_2$, the differential dipole moment $\Delta \bar{d}=(d_1-d_2)/D$, normalized to the molecular dipole moment, and the differential CPV sensitivities.  For ease of comparison, we typically display normalized CPV sensitivities, for example $\bar{P}_d$, defined to be $P_d$ divided by the value to which $P_d$ saturates in the limit of full mixing of the parity doublets.  Note that these values can be larger than 1 since there can be local maxima in sensitivity, as has been observed elsewhere~\cite{Petrov2022YbOHEField,Anderegg2023CaOHSpin}, and because the transitions are between two states, which can have additive sensitivity greater than that of a single state.   There are more than 250 other CP-FITs in $^{173}$YbOH where both $|\Delta \bar{g}|$ and $|\Delta \bar{d}|$ are $<0.01$ and $|\Delta \bar{P}_M|>30 \%$. More CP-FITs can be found when one or two of the three requirements for $|\Delta g|$, $|\Delta \bar{d}|$, and CPV sensitivity are relaxed.
There are also transitions with large field sensitivity and small CPV sensitivity, providing further opportunities for robust systematic error detection and rejection.  Furthermore, these transitions have different relative sensitivities to the eEDM, NSM, and MQM, which is important for disentangling their observable effects. 

\begingroup
\begin{table}
\renewcommand{\arraystretch}{1.3}
\centering
\caption{\label{tab:ybohmagic}Example CP-FITs in $^{173}$YbOH.  The top eight transitions have good CPV sensitivity and low field sensitivity (CP-FITs).  The bottom two transitions have either large electric or magnetic field sensitivity, but small CPV sensitivity.  Note that our uncertainty on $\Delta g$ is $\sim 10^{-3}$}
\begin{ruledtabular}
\begin{tabularx}{1\textwidth}{ccccccc}
$\E$ & $f$ & $|\Delta g|$ & $|\Delta\bar{d}|$ & $|\Delta\bar{P}_d|$ & $|\Delta\bar{P}_Q|$ & $|\Delta\bar{P}_M|$ \\
V/cm & MHz & & & \% & \% & \% \\ \hline
  194.98 &  58 & $1\!\times\!10^{-3}$ &  0 &  24  & 39 &   40 \\ 
  194.48 &  58 & 0 &  $3\!\times\!10^{-4}$ &  24  & 39 &   40 \\ 
  66.71  &  243 & $2\!\times\!10^{-3}$  &  0 &  31 &  42 &  42 \\
  66.55  &  243 & 0  &  $5\!\times\!10^{-5}$ &  31 &  42 &  42 \\ 
  143.1  &  5,132 & $3\!\times\!10^{-3}$  & 0 &  112 &  117 & 141 \\
  238.1 &  23,754 & $3\!\times\!10^{-4}$ &  0 &  93  & 48 &   95 \\ 
  533.91 &  29,520 & $2\!\times\!10^{-4}$ &  0 &  56  & 86 &   84 \\ 
  533.97 &  29,520 & 0 &  $2\!\times\!10^{-4}$ &  56  & 86 &   84 \\ \hline
23.3  &  5,597 &  1.1 & 0  &   1.0  &  0.4  &  0.1 \\
650.6  & 28,550 & 0  &  0.7  &  0.1 & 0.2 &   0.2 \\
\end{tabularx}
\end{ruledtabular}
\end{table}
\endgroup

To quantify the reduction of electromagnetic sensitivity, we calculate decoherence times $\tau_\mathrm{EM}$ due to inhomogeneous field fluctuations $\delta \E \sim$1 mV/cm and $\delta \B \sim$ 1 $\mu$G across the molecular sample, which are typical experimental values~\cite{Baron2017}.  Note the experimental coherence time may be limited to smaller values by several factors, including radiative decay of the bending state, which is estimated to be around 800~ms~\cite{Augenbraun2021YbOCH3}.  An example of a CP-FIT shown in Fig. \ref{fig:magic173result} has $\tau_\mathrm{EM}\sim$2900~s at $\sim$66.7 V/cm.  Both $\Delta g$ and $\Delta \bar{d}$ are suppressed to below 0.001, and therefore have $\tau_\mathrm{EM}\gtrsim$4~s over a 1.2~V/cm range of applied field.  Typical, non-field-insensitive transitions have $|\Delta g|,|\Delta \bar{d}|\sim\mathcal{O}(1)$, which would result in $\tau_\mathrm{EM}\lesssim$~1~ms.  At the same time, the magnitude of the normalized CPV sensitivity maintains reasonably large values of $|\Delta \bar{P}_M|\approx 42 \%$, $|\Delta \bar{P}_d|$ $\approx 31 \%$, and $|\Delta \bar{P}_Q|\approx 42 \%$.

Since there is field dependence of $\Delta g$ and $\Delta \bar{d}$, there will be residual electromagnetic sensitivity in the presence of any field imperfections or noise, though it is also suppressed.  We find $(\text{d}\Delta g/\text{d}\E)\sim2\times 10^{-4}$/(V/cm) and $(\text{d}\Delta \bar{d}/\text{d}\E)\sim1\times 10^{-3}$/(V/cm), indicating that high suppression of $\Delta g$ and $\Delta \bar{d}$ is maintained over more than a few V/cm electric field range and therefore the CP-FIT is robust against the fields gradients and fluctuations.   The dependencies of $\Delta g$ and $\Delta \bar{d}$ on the electric (magnetic) field  are linear over a few V/cm (Gauss) around the zero crossing point, so the differential energy shifts are mostly quadratic.  Note that this dependence is factored in when calculating $\tau_\mathrm{EM}$.

An applied magnetic field parallel (or anti-parallel) to the electric field can be also utilized to further tune $\Delta g$ and $\Delta \bar{d}$. For instance, with a CP-FIT where $\Delta \bar{d}$ crosses zero but $\Delta g$ does not, $\Delta g$ can further be suppressed or even tuned to zero. An example of this exists at electric field of $\sim$143 V/cm and magnetic field of $\sim$106~mG in $^{173}$YbOH, where both the electric and magnetic sensitives vanish. Note that by applying a magnetic field, the exact cancellation of $\Delta \bar{d}$ and the frequency between time-reversed transitions is lost; however, $\Delta \bar{d}$ only changes by $\sim$1$\times 10^{-5}$. Although here we only consider magnetic fields parallel (or anti-parallel) to the electric field, applying a magnetic field at a different angle relative to the electric field, or even to the polarization of an optical dipole trap~\cite{Kotochigova2010Magic,Burchesky2021Rotational}, may provide further tuning of field sensitivities.

While CP-FITs exist quite generically, their location and properties depend strongly on the molecular structure.  To understand these transitions in $^{173}$YbOH, we performed high-resolution spectroscopy on the $\tilde{X}^2\Sigma^+(01^10)$ bending mode in $^{173}$YbOH and $^{171}$YbOH to accurately determine the molecular constants.  We studied the $\tilde{A}^2\Pi_{1/2}(000)-\tilde{X}^2\Sigma^+(01^10)$ band with the same experimental technique and apparatus employed in $^{174}$YbOH recently~\cite{Jadbabaie2023YbOH010}. 
Briefly, a cryogenic buffer gas beam is used to perform laser-induced fluorescence spectroscopy under electric and magnetic fields. The chemical enhancement enables identification of each isotopologue's spectral features~\cite{Pilgram2021YbOHOdd}.  We fit the features to an effective Hamiltonian including Stark, Zeeman, rotation, spin-rotation, $\ell$-doubling, and hyperfine terms, which are shown in table~\ref{tab:params}.  Further details can be found in the Supplemental Material.  

\begingroup
\begin{table}
\renewcommand{\arraystretch}{1.3}
\caption{\label{tab:params}
Measured spectroscopic parameters for the $\tilde{X}^2\Sigma^+(01^10)$ state of YbOH.}
\begin{ruledtabular}
\begin{tabularx}{1\textwidth}{clcc}
Parameter & &	 $^{171}$YbOH &	 $^{173}$YbOH \\ \hline 
$T_0$/cm$^{-1}$ & Origin energy & 319.90861(6) & 319.90932(6)   \\
$B$/MHz & Rotation & 7340.9(3) & 7334.1(4) \\
$\gamma$/MHz & Spin-rotation & $-$90(2) & $-$87(3)  \\
$\gamma_G$/MHz & Axial spin-rotation & 12(6) & 14(4)  \\ 
$q_G$/MHz & $\ell$-doubling & $-$12.6(3) & $-$12.5(5)   \\
$p_G$/MHz & P-odd $\ell$-doubling & $-$11(3) & $-$13(5)  \\
$b_F$/MHz & Fermi contact & 6795(3) & $-$1881.0(8)  \\
$c$/MHz & Spin dipolar & 281(21) & $-$92(10) \\
$e^2Qq_0$/MHz & Quadrupole & \multicolumn{1}{c}{$-$} & -3322(27) \\
\end{tabularx}
\end{ruledtabular}
\end{table}
\endgroup

We varied the spectroscopic parameters in the effective Hamiltonian and confirmed that the CP-FITs still exist with different values of parameters within the experimental uncertainties, with only small changes in their location and properties, as shown in Fig. \ref{fig:magic173result}.  This indicates that the CP-FITs of a particular species can be understood with only moderate uncertainty on spectroscopic parameters, and exist quite generally without reliance on finely-tuned properties.

We then calculated CP-FITs in several species which are candidates for CPV searches and whose spectroscopic parameters have been measured or calculated, such as $^{137}$BaOH~\cite{denis_enhanced_2020, Fletcher1995, Anderson_The_millimeter_BaOH_1993}, $^{173}$YbF~\cite{Ho_YbF_MQM_2023, Wang2019, Sauer1996}, $^{171}$YbF~\cite{Glassman2014, Sauer1996}, $^{171}$YbOH (measured here), $^{174}$YbOH~\cite{Jadbabaie2023YbOH010}, $^{88}$SrOH~\cite{Fletcher1995, steimle_1992_supersonic}, $^{175}$LuOH$^+$~\cite{Maison_TPodd_2022, Alexander_LuOHconstants}, $^{225}$RaF~\cite{Siliviu_RaFconstants, Skripnikov_RaF_2020}, and $^{232}$ThF$^+$~\cite{Gresh_ThF_2016, Kia_Boon_spectroscopy_ThF_2022, Cairncross_thesis}, as shown in table \ref{tab:magic}. We find CP-FITs in all of these species, indicating that they appear to exist quite generally.  However, for species without parity doublets, the electric fields required are orders of magnitude higher than those with parity doublets; this is expected as appreciable mixing of the rotational levels must occur for appreciable CPV sensitivity.

\begingroup
\begin{table}
\renewcommand{\arraystretch}{1.3}
\centering
\caption{\label{tab:magic}Examples of CP-FITs in various species.  See Supplemental Material for details.}
\begin{ruledtabular}
\begin{tabularx}{1\textwidth}{cccccc}
Species & $I$ & $\E$/(V/cm) & $|\Delta \bar{P}_{CPV}|/\%$ & $|\Delta g|$ & $|\Delta\bar{d}|$ \\ \hline 
\multicolumn{3}{c}{$^2\Sigma^+(01^10)$ states} \\
$^{137}$BaOH & \sfrac{3}{2}  & 45.76 & 173 (MQM)  & 9$\times 10^{-4}$ & 0 \\
 &   & 45.74 & 173 (MQM)  & 0 & 9$\times 10^{-4}$ \\
$^{171}$YbOH & \sfrac{1}{2}  & 226.3 & 81 (NSM)  & 2$\times 10^{-3}$ & 0 \\
$^{88}$SrOH & 0  & 223.10 & 94 (eEDM)  & 1$\times 10^{-2}$ & 0 \\
 &  & 223.12 & 94 (eEDM)  & 0 & 4$\times 10^{-3}$ \\
$^{174}$YbOH & 0  & 57.9 & 136 (eEDM)  & 5$\times 10^{-2}$ & 0 \\
$^{175}$LuOH$^+$ & \sfrac{7}{2}  & 519 & 98 (MQM)  & 3$\times 10^{-3}$ & 0 \\ \hline
\multicolumn{3}{c}{$^2\Sigma^+(v=0)$ states} \\
$^{173}$YbF & \sfrac{5}{2}  & 29,976 & 33 (MQM)  & 4$\times 10^{-2}$ & 0 \\
$^{171}$YbF & \sfrac{1}{2}  & 28,170 & 27 (NSM)  & 0 & 5$\times 10^{-6}$ \\
$^{225}$RaF & \sfrac{1}{2}  & 10,215 & 44 (NSM)  & 1$\times 10^{-1}$ & 0 \\
$^{225}$RaF & \sfrac{1}{2}  & 24,568 & 43 (NSM)  & 5$\times 10^{-3}$ & 0 \\
 &   & 24,576 & 43 (NSM)  & 0 & 6$\times 10^{-3}$ \\ \hline
\multicolumn{3}{c}{$^3\Delta_1(v=0)$ states} \\
$^{232}$ThF$^+$ & 0   & 18.6 & 56 (eEDM)  & 1$\times 10^{-4}$ & 0 \\
\end{tabularx}
\end{ruledtabular}
\end{table}
\endgroup

Implementing a CPV measurement utilizing these CP-FITs has the advantage of being immediately compatible with Ramsey methods used, for example, by ACME~\cite{ACME2018} and demonstrated in CaOH ~\cite{Anderegg2023CaOHSpin}, which use either coherent population trapping or microwave pulses, respectively, to create and read out the necessary superpositions. There are many CP-FITs having $\Delta M\leq 2$ (which is a restriction we impose on the CP-FITs considered here), and the transition dipole moment of the two CP-FITs shown in Fig.~\ref{fig:magic173result}, for instance, is $\sim$0.2~Debye, meaning that it can be easily driven with the aforementioned methods.  Reversing the signs of the applied fields, polarizations, etc. can be used to isolate the CPV and non-CPV signals.  The same measurement could be performed on the time-reversed pair to make use of the internal co-magnetometer feature; simultaneous measurement of both transitions could provide very strong robustness.

While the ability to suppress the effects of electromagnetic fields is valuable, any CPV measurement requires a careful and thorough consideration of systematic effects, which will be unique to each system.  Here we mention two example relevant effects. 

First, the residual, quadratic field sensitivity mentioned earlier will result in false EDMs due to non-reversing (nr) background fields~\cite{Baron2017}, which can be estimated to be $\{\mu_B \frac{\text{d}\Delta g}{\text{d}\E} + D  \frac{\text{d}\Delta \bar{d}}{\text{d}\B} \} \E_{nr} \B_{nr} \sim$0.3 $\mu$Hz, or $\lesssim10^{-31}$~e~cm, using $\E_{nr}=1~$~mV/cm and $ \B_{nr}=$1~$\mu$G, and can be suppressed by using the internal co-magnetometers.  Furthermore, there are different CP-FITs where $\frac{\text{d}\Delta g}{\text{d}\E}$/$\Delta P_{CPV}$ differs by a factor of $\sim$100. This means that the false vs. true CPV signals will be different for these transitions.  Note that molecules in optical traps will have tensor AC Stark residual moments as well, as was observed in CaOH~\cite{Anderegg2023CaOHSpin}, which could be addressed using the same methods discussed therein.

Second, transverse magnetic fields will introduce coupling between different $M_F$ states and cause unwanted precession and dephasing.  These states are typically split by $\delta \gtrsim$100~kHz due to fine and hyperfine structure, so a transverse field of $\sim\mu$G will give rise to a transverse coupling of $\Omega \sim$1 Hz and second order shift on the order of $\frac{\Omega^2}{4\delta} \sim \frac{(1  \text{Hz})^2}{4(100  \text{kHz})} \sim\mu$Hz.  Transitions involving $|M_F| =$1/2 states will have a degenerate partner state directly coupled by a transverse field; such a degeneracy could be lifted by applying a small bias field, with $\sim$1~mG giving rise to a second order shift of $\lesssim$mHz.

In summary, we find transitions in molecules with electric and magnetic field sensitivity suppressed by over a hundred-fold while maintaining good sensitivity to CP-violation.  These transitions offer internal co-magnetometry, and can be driven with RF, microwave, or two-photon transitions, and are therefore compatible with Ramsey measurements in both beams and traps. It is also suitable for systems with large angular momentum commonly found in nuclear CPV searches where hyperfine structure increases the internal complexity and congestion.  No particular molecular structure is required, though in species with parity doublets the transitions exist at moderate electric fields $\lesssim$1~kV/cm.  It is likely that similar transitions exist in a wide range of species, including atoms~\cite{Yb_MQM_paper}, and both symmetric and asymmetric tops. They may also be compatible with spin-squeezing methods to decrease the noise limit.  Finally, the ability to polarize molecules while suppressing decoherence from electromagnetic field noise with these transitions can be of great interest in other fields including quantum simulation, where a long-range dipole-dipole interaction with a long interaction time is extremely advantageous~\cite{Gambetta_engineering_2020, Gambetta_long-range_2020, Gambetta_exploring_2021}. 

\begin{acknowledgments}

We acknowledge helpful discussions with Chris Ho, Zack Lasner, Tim Steimle, Amar Vutha, and Trevor Wright.  We thank Alexander Petrov, Ronald Garcia Ruiz, and Silviu-Marian Udrescu for providing molecular constants for our calculations.  This work was supported by the Heising-Simons Foundation (2019-1193 and 2022-3361), the Alfred P. Sloan Foundation (G-2019-12502), the Gordon and Betty Moore Foundation (GBMF7947), and an NSF CAREER Award (PHY-1847550). Y. T. was supported by the Masason Foundation. C. Z. was supported by the David and Ellen Lee Postdoctoral Scholarship.

\end{acknowledgments}

\bibliography{biblio,ref2}

\newpage
\onecolumngrid
\setcounter{section}{0}
\setcounter{equation}{0}
\setcounter{figure}{0}
\setcounter{table}{0}
\makeatletter

\newpage

\input{supplemental}

\end{document}

%% file: supplemental.tex
\section*{Supplemental Material: Engineering field-insensitive molecular clock\\ transitions for symmetry violation searches}

\FloatBarrier

\subsection{Examples of CP-FITs in various species}

\begingroup
\begin{table*}
\renewcommand{\arraystretch}{1.0}
\centering
\begin{threeparttable}
\caption{\label{tab:otherspecies}Examples of CP-FITs in other species. }
\begin{ruledtabular}
\begin{tabularx}{1\textwidth}{cccccccccccc}

Species & $I$ & State/Transition$^a$ & Frequency & $M_{F^{\prime\prime}}, M_{F^{\prime}}$ & E field & $|\bar{P}_d|$$^b$ & $|\bar{P}_Q|$$^b$ & $|\bar{P}_M|$$^b$  &  $|\Delta \bar{d}|$ &  $|\Delta g|$ & $\tau_\textrm{EM}$ \\ 
& & & MHz & & $[$V/cm$]$ &   \% & \% & \% & & & sec\\ \hline 
\multicolumn{3}{c}{$^2\Sigma^+(01^10)$ states} \\
$^{137}$BaOH$^c$ & 3/2 & $N = 1$ & 21.0 & 3/2, 3/2 & 45.76 & 69 & 117 & 173 & zero crossing & 9.0$\times 10^{-4}$ & 729 \\
& & & &  & 45.74 & 69 & 117 & 173 & 9.3$\times 10^{-4}$ & zero crossing & 1.0 \\
$^{171}$YbOH & 1/2 & $N = 1$ & 6.8 & 3/2, 3/2 & 226.3 & 81& 81 &  \multicolumn{1}{c}{$-$} & zero crossing &  1.6$\times 10^{-3}$ & 435 \\
$^{88}$SrOH$^d$ & 0 & $N = 2$ & 120.46 & 2, 1 & 223.10 & 94 &  \multicolumn{1}{c}{$-$} &  \multicolumn{1}{c}{$-$} & zero crossing & 1.0$\times 10^{-2}$ & 70 \\
& & & &  & 223.12 & 94 &  \multicolumn{1}{c}{$-$} &  \multicolumn{1}{c}{$-$} & 3.8$\times 10^{-3}$ & zero crossing & 0.2 \\
$^{174}$YbOH & 0 & $N=1\rightarrow 2$ & 29196 & 1, 1 & 57.9 & 136 &  \multicolumn{1}{c}{$-$} &  \multicolumn{1}{c}{$-$} & zero crossing & 5.4$\times 10^{-2}$  & 13 \\
$^{175}$LuOH$^+$ & 7/2 & $N=1$ & 32895 & 1/2, --3/2 & 519 & 38 &  151  &  98 & zero crossing & 3.3$\times 10^{-3}$  & 216 \\ \hline
\multicolumn{3}{c}{$^2\Sigma^+(v=0)$ states} \\
$^{173}$YbF$^e$ & 5/2 & $N=1$ & 5375.9 & 1/2, --3/2 & 29973 & 18 & 41 & 33 & zero crossing & 3.9$\times 10^{-2}$ & 18.1 \\
$^{171}$YbF$^f$ & 1/2 & $N=0$ & 7312.8 & 1/2, --3/2 & 28170 & 27 & 27 &  \multicolumn{1}{c}{$-$} & 5.0$\times 10^{-6}$ & zero crossing  & 185 \\
$^{225}$RaF$^g$ & 1/2 & $N = 1$ & 121.9 & 1/2, 3/2 & 10215 & 44 & 44 &  \multicolumn{1}{c}{$-$} & zero crossing & 9.6$\times 10^{-2}$ & 7.4 \\
$^{225}$RaF$^g$ & 1/2 & $N = 1$ & 215.9 & 1/2, --3/2 & 24569 & 43 & 43 &  \multicolumn{1}{c}{$-$} & zero crossing & 4.8$\times 10^{-3}$ & 148 \\
& & & &  & 24577 & 43 & 43 &  \multicolumn{1}{c}{$-$} & 5.9$\times 10^{-3}$ & zero crossing & 0.2 \\ \hline
\multicolumn{3}{c}{$^3\Delta_1(v=0)$ states} \\
$^{232}$ThF$^{+h}$ & 0 & $J=1\rightarrow 2$ &  29085 & 1/2, 1/2 & 18.6 & 56 &  \multicolumn{1}{c}{$-$} &  \multicolumn{1}{c}{$-$} & zero crossing & 1.1$\times 10^{-4}$ & 6656 \\
\end{tabularx}
\end{ruledtabular}
\begin{tablenotes}
\item{$^a$} For the CP-FITs with relatively large electric fields of $\gtrsim$1 kV/cm, the rotational quantum number shown is that which constitutes the largest admixture as $N$ is no longer a good quantum number.
\item{$^b$} For $^{137}$BaOH, $^{171}$YbOH, $^{88}$SrOH, $^{174}$YbOH, and $^{232}$ThF$^+$, sensitivities are normalized to the value to which $|\bar{P}|$ saturates in the limit of full mixing of the parity doublets. For $^{173}$YbF, $^{171}$YbF, and $^{225}$RaF, sensitivities are normalized to their maximum sensitivity obtained over all $N$ = 0--2 states in 0--50 kV/cm electric field range.
\item{$^c$} Spectroscopic parameters are taken from ref.~\cite{denis_enhanced_2020, Anderson_The_millimeter_BaOH_1993}. Since there is no direct measurement of $^2\Sigma^+(010)$ state, we assumed that the spectroscopic parameters and molecular dipole moment are the same as $^2\Sigma^+(000)$ state. The $q_G$ parameter is assumed to be the same as $^{138}$BaOH~\cite{Fletcher1995}.
\item{$^d$} Parameters are taken from ref.~\cite{Fletcher1995, steimle_supersonic_1992}. The molecular dipole moment is assumed to be the same as $^2\Sigma^+(000)$ state. 
\item{$^e$} Spectroscopic parameters are taken from ref.~\cite{Wang2019, Sauer1996}. The molecular dipole moment is assumed to be the same as $^{174}$YbF.
\item{$^f$} Spectroscopic parameters are taken from ref.~\cite{Glassman2014, Sauer1996}. The molecular dipole moment is assumed to be the same as $^{174}$YbF.
\item{$^g$} Spectroscopic parameters are taken from ref.~\cite{Siliviu_RaFconstants}. The hyperfine parameters from fluorine and molecular dipole moment are taken from the theoretical calculation results~\cite{Skripnikov_RaF_2020}.
\item{$^h$} Spectroscopic parameters are taken from ref.~\cite{Gresh_ThF_2016, Kia_Boon_spectroscopy_ThF_2022}. The effective Hamiltonian and its matrix element in  $^3\Delta_1$ state are taken from ref.~\cite{Cairncross_thesis}. Here, we neglected transverse magnetic and electric fields. The g-factor is assumed to be the same for all states as the $J$ = 1, $F$ = 3/2 state g-factor. Because the spin-orbit interaction in $^3\Delta_1$ state couples $\Sigma$ and $\Lambda$ and aligns them in anti-parallel, the orbital and electron magnetic moments nearly cancel each other, offering highly suppressed g-factor~\cite{Hutzler2014Thesis}. Thus, $\Delta g$ is generally small in most of the transitions. This is in contrast with $^2\Sigma^+$ state where there is no cancellation due to the spin-orbit interaction. Note also again that the contributions from the nuclear and rotational zeeman terms are neglected.
\item{$^i$} For ion species (LuOH$^+$ and ThF$^+$), transverse magnetic and electric fields in the rotating frame are neglected.
\item{$^j$} Hyperfine constants from hydrogen nuclear spin are assumed to be the same as $^{174}$YbOH.
\item{$^k$} For all species except YbOH and ThF$^+$, the effective g-factor is assumed to be the same as the electron g-factor $g_S$.
\end{tablenotes}

\end{threeparttable}
\end{table*}
\endgroup

\FloatBarrier
\subsection{Calculation of coherence time}

The frequency of the transition is shifted by the electric and magnetic field in the following way:


\begin{subequations}
\begin{align}
    & f = D \Delta \bar{d}\cdot \E +  \mu_B \Delta g\cdot \B
\end{align}
\end{subequations}

The broadening of the transition due to inhomogeneous electric and magnetic field fluctuation is thus given by:

\begin{subequations}
\begin{align}
    & \delta f_{\delta \E} = D \Bigl\{\Delta \bar{d}\cdot \delta \E + \frac{\text{d}\Delta \bar{d}}{\text{d}\E}\cdot (\delta \E)^2 + \frac{\text{d}\Delta \bar{d}}{\text{d}\B}\cdot \delta B \cdot \delta \E\Bigr\} \\
    & \delta f_{\delta \B} = \mu_B \Bigl\{\Delta g\cdot \delta \B +  \frac{\text{d}\Delta g}{\text{d}\E}\cdot \delta \E \cdot \delta \B + \frac{\text{d}\Delta g}{\text{d}\B}\cdot (\delta \B)^2\Bigr\}
\end{align}
\end{subequations}

The electromagnetic decoherence time $\tau_\textrm{EM}$ is thus given by:

\begin{subequations}
\begin{align}
    & \tau_\textrm{EM} =  \frac{1}{\delta f_{\delta \E} + \delta f_{\delta \B}}
\end{align}
\end{subequations}

\subsection{Lifting the degeneracy in $M_F=\pm 1/2$ states}

$|M_F| =$1/2 states are degenerate in the absence of a $B_z$ field and can be coupled by transverse fields at first order, whereas $|M_F| \neq$1/2 states are typically split by more than 100 kHz due to fine and hyperfine structure and thus at second or higher order. 
The application of a small bias field would lift the degeneracy of $|M_F| =$1/2 states to suppress the coupling between the states. 
For instance, with a bias field of 1 mG, $M_F=\pm 1/2$ states are split by $\sim$kHz, and therefore there is a quadratic sensitivity to transverse fields of $\ll$~mG.   This will break the perfect symmetry of the internal co-magnetometry, but by a small amount; for example in the transition shown in figure 2 in $^{173}$YbOH, $\Delta \bar{d}$, $\Delta g$, and the transition frequencies change by $\sim$1$\times 10^{-6}$, $\sim$3$\times 10^{-5}$, and $\sim$Hz respectively. The frequency difference between the \textit{left} and \textit{right} transition can be exactly canceled out by reversing the $E_z$ or $B_z$ field individually.

\subsection{Effective Hamiltonian}

In $^{173}$YbOH, the $^{173}$Yb nucleus has a nuclear spin of 5/2, and the hyperfine coupling between $^{173}$Yb nuclear spin and an electron spin forms the angular momentum $\bm{G} = \bm{I} + \bm{S}$. (Here, $\bm{S}$ and $\bm{I}$ are the electron and nuclear spin of Yb respectively.) $\bm{G}$ is then coupled to the rotational angular momentum $\bm{N}$, forming the angular momentum $\bm{F_1} = \bm{G} + \bm{N}$. Here, $\bm{N} = \bm{\Lambda} + \bm{\ell} + \bm{R}$ where $\bm{R}$ is the nuclei rotational angular momentum, $\bm{\Lambda}$ is the projection of electronic orbital angular momentum on the molecular axis, and $\bm{\ell}$ is the projection of vibrational angular momentum on the molecular axis.
Finally, the total angular momentum $\bm{F}$ is formed from the coupling between $\bm{F_1}$ and $\bm{I_H}$, where $\bm{I_H}$ is the distant Hydrogen nuclear spin. The non Born-Oppenheimer interactions give rise to parity doublets of $|\mathcal{P}=\pm\ket =\frac{1}{\sqrt{2}} (|N,\ell\ket \pm (-1)^{N-\ell-\Lambda} |N,-\ell\ket)$.

The total effective molecular Hamiltonian for $\tilde{X}^2\Sigma^+(01^10)$ state of $^{171, 173}$YbOH used in the numerical diagonalization is given by~\cite{Pilgram2021YbOHOdd}:

\begin{subequations}
\begin{align}
    & H_{\tilde{X}(010)} = H_{Rot} + H_{SR} + H_{aSR} + H_{b_F(\text{Yb})}  +  H_{c(\text{Yb})}+ H_{b_F(\text{H})} +  H_{c(\text{H})}  + H_{Q} + H_{q_G} + H_{p_G}  + H_{Stark} + H_{Zeeman},
\end{align}
\end{subequations}
where
\begin{subequations}
\begin{align}
    & H_{Rot} = B (N^2-\ell^2)  \\
    & H_{SR} = \gamma (N\cdot S - N_z S_z) \\
    & H_{aSR} = \gamma_{G} N_z S_z  \\
    & H_{b_F(\text{Yb})} = b_F(^{171,173}\text{Yb}) I\cdot S \\
    & H_{c(\text{Yb})} = c(^{171,173}\text{Yb}) (I_z S_z - \frac{1}{3} I\cdot S) \\
    & H_{b_F(\text{H})} = b_F(\text{H}) I\cdot S \\
    & H_{c(\text{H})} = c(\text{H}) (I_z S_z - \frac{1}{3} I\cdot S) \\    
    & H_{Q} = e^2Qq_0(^{173}\text{Yb}) \frac{3I_z^2 - I^2}{4I(2I-1)} \\
    & H_{q_G} = - \frac{q_G}{2} (N_+^2 e^{-i 2 \phi} + N_-^2 e^{i2 \phi})  \\
    & H_{p_G} =  \frac{p_G}{2} (N_+S_+ e^{-i 2 \phi} + N_-S_- e^{i2 \phi})  \\
    & H_{Stark} = -D_\mathrm{\tilde{X}}\cdot E \\
    & H_{Zeeman} = g_S \mu_B S_Z B_Z.
\end{align}
\end{subequations}
$\phi$ is the azimuthal angle of the bending nuclear framework, $D_\mathrm{\tilde{X}}$ is the dipole moment of $\tilde{X}^2\Sigma^+(01^10)$ state in the molecule frame, +/-- and z subscripts refer to the molecule frame projections, and $Z$ refers to the lab-frame projection.  The total Hamiltonian consists of rotation ($H_{Rot}$), spin-rotation ($H_{SR}$), axial spin-rotation ($H_{aSR}$), Fermi contact (Yb) ($H_{b_F(\text{Yb})}$), dipolar (Yb) ($H_{c(\text{Yb})}$), Fermi contact (Hydrogen) ($H_{b_F(\text{H})}$), dipolar (Hydrogen) ($H_{c(\text{H})}$), quadrupole ($H_{Q}$), $\ell$-doubling $q_G$ ($H_{q_G}$), $\ell$-doubling $p_G$ ($H_{p_G}$), Stark ($H_{Stark}$), and Zeeman ($H_{Zeeman}$) terms. The quadrupole term only exists for $^{173}$YbOH. Subtraction of $N_z S_z$ in spin-rotation term and sign convention for $\ell$-doubling term are detailed in ref.~\cite{Jadbabaie2023YbOH010}.

$B$, $\gamma$, $\gamma_{G}$, $b_F$(Yb), $c$(Yb), $e^2Qq_0$, $q_G$, and $p_G$ are spectroscopic constants for each term and determined by spectroscopy on $\tilde{A}^2\Pi_{1/2}(000)-\tilde{X}^2\Sigma^+(01^10)$ band. $b_F$(H) and $c$(H) are measured to be 0.000160 and 0.000082 cm$^{-1}$ in the previous microwave study in $^{174}$YbOH~\cite{Jadbabaie2023YbOH010}. $D_\mathrm{\tilde{X}}$ and $g_S$ are also measured to be 2.16 D and 2.07 respectively in $^{174}$YbOH.

The Hamiltonian terms above are represented in the spherical tensor notation in the following way:

\begin{subequations}
\begin{align}
    & H_{Rot} = B (N^2-\ell^2)  \\
    & H_{SR} = \gamma \left(T^1(N)\cdot T^1(S) - T^1_{q=0}(N) T^1_{q=0}(S)\right) \\
    & H_{aSR} = \gamma_G T^1_{q=0}(N) T^1_{q=0}(S)  \\
    & H_{b_F(\text{Yb})} = b_F(^{171,173}\text{Yb}) T^1(I)\cdot T^1(S)  \\
    & H_{c(\text{Yb})} = c(^{171,173}\text{Yb}) \frac{\sqrt{6}}{3} T^2_{q=0}(I, S) \\
    & H_{b_F(\text{H})} = b_F(\text{H}) T^1(I)\cdot T^1(S)  \\
    & H_{c(\text{H})} = c(\text{H}) \frac{\sqrt{6}}{3} T^2_{q=0}(I, S) \\    
    & H_{Q} = e^2Qq_0(^{173}\text{Yb}) \frac{\sqrt{6}}{4I(2I-1)} T^2_{q=0}(I, I) \\
    & H_{q_G} = - q_G \sum_{q=\pm1} e^{-2iq \phi} T^2_{2q}(N,N) \\
    & H_{p_G} =  p_G \sum_{q=\pm1} e^{-2iq \phi} T^2_{2q}(N,S) \\
    & H_{Stark} = - T^1_{p=0}(D_\mathrm{\tilde{X}}) T^1_{p=0}(E)  \\
    & H_{Zeeman} = g_S \mu_B T^1_{p=0}(S) T^1_{p=0}(B)
\end{align}
\end{subequations}

Note that $p(q)$ labels space-fixed (molecule-fixed) spherical tensor coordinates.

\subsection{Spectroscopy of $\tilde{X}^2\Sigma^+(01^10)$ state in $^{171,173}$YbOH}

As mentioned in the main text, we used the same experimental technique and apparatus employed in $^{174}$YbOH to conduct high-resolution spectroscopy on $\tilde{A}^2\Pi_{1/2}(000)-\tilde{X}^2\Sigma^+(01^10)$ band in $^{171, 173}$YbOH~\cite{Jadbabaie2023YbOH010}. 

Briefly, a YbOH molecular beam from the cryogenic buffer-gas source enters a laser-induced fluorescence (LIF) detection region where the molecules can experience electric and magnetic fields from ITO-coated glass electrodes and magnetic field coils, respectively. By utilizing the isotopolgue specific enhancement technique $^{173}$YbOH spectra are disentangled from different isotopologues even in spectral regions where they are significantly overlapped. 

The effective molecular Hamiltonian for the $\tilde{A}^2\Pi_{1/2}(000)$ state used to model the spectrum without Stark and Zeeman terms is detailed in ref.~\cite{Pilgram2021YbOHOdd}. It includes spin-orbit ($A_{SO}$), rotation ($B$), centrifugal distortion ($D$), $\Lambda$-doubling ($p+2q$), quadrupole ($e^2Qq_0$), and hyperfine a and d terms ($a$, $d$).

The Stark and Zeeman terms for the $\tilde{A}^2\Pi_{1/2}(000)$ state is given by:
\begin{equation}
\begin{split}
    & H_{Stark} = -D_\mathrm{\tilde{A}}\cdot E \\
    & H_{Zeeman} = g^\prime_S \mu_B S_Z B_Z + g_L L_Z B_Z + g^\prime_l \mu_B (e^{-2i\theta} S_+ B_+ + e^{2i\theta} S_- B_-), 
\end{split}
\end{equation}
where $D_\mathrm{\tilde{A}}$ is the dipole moment of $\tilde{A}^2\Pi_{1/2}(000)$ state in the molecule frame, and $\theta$ is the electronic azimuthal coordinate. $D_\mathrm{\tilde{A}}$ has been previously measured to be  0.43 D ~\cite{Steimle2019}. For $g^\prime_S$, $g_L$, and $g^\prime_l$. we use the values from Ref.~\cite{Sawaoka_ZS_YbOH_2023}, fixing $g^\prime_S=1.860$, $g_L=1.0$, and $g^\prime_l = -0.724$.

The effective Hamiltonian for $\tilde{X}^2\Sigma^+(01^10)$ is described in the previous section. Again, it includes rotation ($B$), spin-rotation ($\gamma$), axial spin-rotation ($\gamma_{G}$), Fermi contact (Yb) ($b_F$(Yb)), dipolar (Yb) ($c$(Yb)), quadrupole ($e^2Qq_0$), $\ell$-doubling $q_G$ ($q_G$), and $\ell$-doubling $p_G$ ($p_G$) terms.

The hyperfine structure from the distant Hydrogen nuclear spin $I_H$ is neglected in the spectroscopy analysis since it is optically unresolved~\cite{Jadbabaie2023YbOH010}. However, in the main analysis, we included the hyperfine structure and set $b_F$(H) and $c$(H) to 0.000160 and 0.000082 cm$^{-1}$, which were previously determined for $^{174}$YbOH~\cite{nakhate_pure_2019_no_doi}.

Since the spectroscopic constants $B$, $\gamma$, and $q_G$ of the $\tilde{X}^2\Sigma^+(01^10)$ state in $^{174}$YbOH are measured, those of $^{171,173}$YbOH can be initially guessed with high accuracy using mass scaling. The spectroscopic constants $A_{SO}$, $B$, $D$, $p+2q$, $e^2Qq_0$, $a$, and $d$ of $\tilde{A}^2\Pi_{1/2}(000)$ and $b_F$, $c$, and $e^2Qq_0$ of the $\tilde{X}^2\Sigma^+(000)$ state in $^{171,173}$YbOH are also determined from the previous work~\cite{Pilgram2021YbOHOdd}. By combining all the information, the spectrum can be predicted with high accuracy and directly compared against the observed spectrum to assign the observed lines.

Since this band is forbidden in E1 approximation, the intensities arise from the perturbations in the electronic excited state. The study of $^{174}$YbOH modeled the intensity borrowing and their prediction agrees well with the observed intensities~\cite{Jadbabaie2023YbOH010}. The similar intensity borrowing coefficients of $(c_\mu, c_\kappa, c_B) = (0.20, -0.40, 0.89)$ are used to predict the intensities in $^{171,173}$YbOH to assist the line assignments and they showed good agreement with the observed spectrum. For the lines that are close and overlapped each other, they are assigned to the same feature. As a result, in total, 37 and 61 transitions are assigned in $^{171}$YbOH and $^{173}$YbOH, respectively. 

We obtained the spectroscopic constants by fitting the assigned lines. For the overlapped lines that are assigned to the same feature, they are weighted according to the peak width in the fitting.

As expected from the mass scaling, the rotational constant decreases as the mass of the Yb nucleus increases. The parameters associated with $\ell$-doubling ($q_G$, $p_G$, and $\gamma_G$) are the same within their errors, which are also expected due to the small dependence on the mass. The hyperfine and quadrupole constants ($b_F$, $c$, and $e^2Qq_0$) are the same as or close to the measured values for $\tilde{X}^2\Sigma^+(000)$ within the errors~\cite{Pilgram2021YbOHOdd}. The relatively large errors on $c$ and $e^2Qq_0$ are presumably due to the large errors on the excited $\tilde{A}^2\Pi_{1/2}(000)$ state parameters. Nevertheless, our fit residuals on the assigned lines are smaller than the previous spectroscopy work on $\tilde{A}^2\Pi_{1/2}(000)-\tilde{X}^2\Sigma^+(000)$ band in $^{171, 173}$YbOH by a factor of few~\cite{Pilgram2021YbOHOdd}.

The Stark and Zeeman spectra under the DC electric and magnetic fields were also measured and compared against the prediction with the molecular dipole moment and $g_S$ fixed at 2.16 D and 2.07 respectively, which were measured for $^{174}$YbOH~\cite{Jadbabaie2023YbOH010}. The electric and magnetic fields lift the Zeeman sublevel degeneracy, resulting in the significantly congested spectra. However, both the overall structure and the shift of the strong features agree with the prediction at each field value, confirming that the dipole moment of $^{171,173}$YbOH is roughly the same as $^{174}$YbOH, which is expected.   We do not report a measured value of the dipole moments, but instead use our measurements as a check that our assumption of their similarity between isotopologues to be accurate, and fix the moments at the values measured for $^{174}$YbOH.  A more comprehensive analysis of Stark and Zeeman spectrum would require higher resolution via microwave, RF, or optical Raman transitions to resolve the small splittings between different Zeeman sublevels.


\begingroup
\begin{table*}
\renewcommand{\arraystretch}{1.4}
\centering
\begin{threeparttable}
\caption{\label{tab:params}
Spectroscopic parameters for the $\tilde{X}^2\Sigma^+(01^10)$ state of YbOH.}
\begin{ruledtabular}
\begin{tabularx}{1\textwidth}{llccc}

Parameter &	&  $^{171}$YbOH &	 $^{173}$YbOH &	 $^{174}$YbOH~\cite{Jadbabaie2023YbOH010} \\ \hline 
$T_0$/cm$^{-1}$ & Origin energy & 319.90861(6) & 319.90932(6) & 319.90901(6)  \\
$B$/MHz & Rotation & 7340.9(3) & 7334.1(4) & 7,328.64(15) \\
$\gamma$/MHz & Spin-rotation & --90(2) & --87(3) & --88.7(9)  \\
$\gamma_G$/MHz & Axial spin-rotation & 12(6) & 14(4) & 16(2)  \\ 
$q_G$/MHz & $\ell$-doubling & --12.6(3) & --12.5(5) & --12.0(2)  \\
$p_G$/MHz & P-odd $\ell$-doubling & --11(3) & --13(5) & --11(1) \\
$b_F$/MHz & Fermi contact & 6795(3) & --1881.0(8)  & \multicolumn{1}{c}{$-$} \\
$c$/MHz & Spin dipolar & 281(21) & --92(10) & \multicolumn{1}{c}{$-$} \\
$e^2Qq_0$/MHz & Quadrupolar & \multicolumn{1}{c}{$-$} & --3322(27) & \multicolumn{1}{c}{$-$} \\
\end{tabularx}
\end{ruledtabular}
\end{threeparttable}
\end{table*}
\endgroup

\subsection{Measurement errors of spectroscopic parameters and their impact on CP-FITs}

To understand the effect of spectroscopic parameter variation on the position and qualities of the CP-FITS, we varied each individual spectroscopic parameter of $^{173}$YbOH by its fit error. Many of the parameters do not have a significant impact on the quality of the CP-FITs and they simply slightly shift the electric field value for the CP-FIT. Despite the relatively large errors on $b_F$ and $c$, they only shift $\Delta \bar{d}$ slightly by $\ll$0.001. $e^2Qq_0$ has also large error of $\sim$27 MHz and $q_G$ and $p_G$ are the parameters sensitive to the parity doublet structure. Nonetheless, variation of $e^2Qq_0$ and $p_G$ only give rise to the uncertainty of $\Delta \bar{d}$ of $<$0.001. Variation of $q_G$ has the biggest impact on the CP-FIT, leading to the uncertainty of $\Delta \bar{d}$ and $\Delta g$ on the order of $\sim$0.002 and $\sim$0.0005 respectively. Note that $\Delta g$ $\sim$0.0005 is comparable to or below the expected contribution of $\Delta g$ $\sim$0.001 from nuclear spin and rotational Zeeman terms.


Combined all above, it is fair to claim that optical spectroscopy would be sufficient to identify the candidates of CP-FITs and predict their qualities with reasonably good accuracy. Their qualities can be confirmed in the subsequent experiment with higher precision such as microwave, RF, or two-photon Raman.

\subsection{Matrix elements}

The matrix elements are evaluated in the following case (b) basis and used to calculate the structure and CPV sensitivities. Note that $K=\Lambda + \ell = N\cdot \hat{n}$. In $\tilde{X}^2\Sigma^+(01^10)$ state, $\Lambda$ = 0.

The rotation term:

\begin{equation}
    \begin{split}
      \langle K; N, S, I, G, F_1, I_H, F, M |& (N^2-\ell^2)| K^\prime; N^\prime, S, I, G^\prime, F_1^\prime, I_H, F^\prime, M^\prime  \rangle \\
         &= \delta_{F,F^\prime} \delta_{M,M^\prime} \delta_{F_1,F_1^\prime} \delta_{G,G^\prime} \delta_{N, N^\prime} \delta_{K,K^\prime}\lbrace N(N+1) - K^2 \rbrace
    \end{split}
\end{equation}

The spin-rotation term (without subtraction of axial spin-rotation term):

\begin{equation}
    \begin{split}
     \langle K; N, S, I, G, F_1, I_H, F, M |& T^1(N) \cdot T^1(S)| K^\prime; N^\prime, S, I, G^\prime, F_1^\prime, I_H, F^\prime, M^\prime  \rangle \\
         &= \delta_{F,F^\prime} \delta_{M,M^\prime} \delta_{F_1,F_1^\prime} \delta_{K,K^\prime} \delta_{N,N^\prime} \\
         &  \times (-1)^{N^\prime + F_1 + G} \sixJ{G^\prime}{N^\prime}{F_1}{N}{G}{1}   \\
         &  \times \sqrt{N(N+1)(2N+1)}\\
         &  \times (-1)^{G^\prime + S + 1 + I} \sqrt{(2G+1)(2G^\prime+1)} \sixJ{S}{G^\prime}{I}{G}{S}{1} \sqrt{S(S+1)(2S+1)}
    \end{split}
\end{equation}

The axial spin-rotation term:

\begin{equation}
    \begin{split}
     \langle K; N, S, I, G, F_1, I_H, F, M |& T^1_{q=0}(N) T^1_{q=0}(S)| K^\prime; N^\prime, S, I, G^\prime, F_1^\prime, I_H, F^\prime, M^\prime  \rangle \\
         &= \delta_{F,F^\prime} \delta_{M,M^\prime} \delta_{F_1,F_1^\prime} \\
         &  \times (-1)^{N^\prime + F_1+ G} \sixJ{G^\prime}{N^\prime}{F_1}{N}{G}{1}   \\
         &  \times K \times  (-1)^{N-K} \threeJ{N}{-K}{1}{0}{N^\prime}{K^\prime} \sqrt{(2N+1)(2N^\prime+1)}\\
         &  \times (-1)^{G^\prime + S + 1 + I} \sqrt{(2G+1)(2G^\prime+1)} \sixJ{S}{G^\prime}{I}{G}{S}{1} \sqrt{S(S+1)(2S+1)}
    \end{split}
\end{equation}

This can be derived formally by using eq. (2.3.35) in ref.~\cite{hirota_high-resolution_1985}. \\ \\

The Fermi contact (Yb) term:

\begin{equation}
    \begin{split}
     \langle K; N, S, I, G, F_1, I_H, F, M |&  T^1(I) \cdot T^1(S) | K^\prime; N^\prime, S, I, G^\prime, F_1^\prime, I_H, F^\prime, M^\prime  \rangle \\
        & = \delta_{F,F^\prime} \delta_{M,M^\prime} \delta_{F_1,F_1^\prime} \delta_{G, G^\prime} \delta_{N, N^\prime} \delta_{K, K^\prime}\\
    &  \times (-1)^{I + G^\prime + S}  \sixJ{S}{I}{G}{I}{S}{1} \sqrt{S(S+1)(2S+1)} \sqrt{I(I+1)(2I+1)}
    \end{split}
\end{equation}

The Dipolar (Yb) term:

\begin{equation}
    \begin{split}
     \langle K; N, S, I, G, F_1, I_H, F, M |&  T^2(I, S) \cdot T^2(n) | K^\prime; N^\prime, S, I, G^\prime, F_1^\prime, I_H, F^\prime, M^\prime  \rangle \\
    & = \delta_{F,F^\prime} \delta_{M,M^\prime} \delta_{F_1,F_1^\prime} \\
    &  \times (-1)^{N^\prime + F_1 + G} \sixJ{G^\prime}{N^\prime}{F_1}{N}{G}{2}   \\
    & \times (-1)^{N-K} \sqrt{(2N+1)(2N^\prime+1)} \threeJ{N}{-K}{2}{0}{N^\prime}{K^\prime} \\
    & \times \sqrt{(2G+1)5(2G^\prime+1)} \nineJ{I}{S}{G^\prime}{1}{1}{2}{I}{S}{G} \\
    & \times  \sqrt{S(S+1)(2S+1)} \sqrt{I(I+1)(2I+1)} \\    
    \end{split}
\end{equation}

The Fermi contact (H) term:

\begin{equation}
    \begin{split}
     \langle K; N, S, I, G, F_1, I_H, F, M |&  T^1(I_H) \cdot T^1(S) | K^\prime; N^\prime, S, I, G^\prime, F_1^\prime, I_H, F^\prime, M^\prime  \rangle \\
     & = \delta_{F,F^\prime} \delta_{M,M^\prime} \delta_{N, N^\prime} \delta_{K, K^\prime} \\
     & \times (-1)^{F_1^\prime+F+I_H} \sixJ{I_H}{F_1^\prime}{F}{F_1}{I_H}{1} \sqrt{I_H(I_H+1)(2I_H+1)} \\
     & \times (-1)^{F_1+N+1+G^\prime} \sqrt{(2F_1+1)(2F_1^\prime+1)} \sixJ{G^\prime}{F_1^\prime}{N}{F_1}{G}{1} \\
    &  \times (-1)^{G + I + 1 + S} \sqrt{(2G+1)(2G^\prime+1)} \sixJ{S}{G^\prime}{I}{G}{S}{1} \sqrt{S(S+1)(2S+1)}
    \end{split}
\end{equation}

The Dipolar (H) term:

Using eq. (2.3.80) in ref.~\cite{hirota_high-resolution_1985},

\begin{equation}
    \begin{split}
     \langle K; N, S, I, G, F_1, I_H, F, M |& T^2_{q=0}(I_H, S) | K^\prime; N^\prime, S, I, G^\prime, F_1^\prime, I_H, F^\prime, M^\prime  \rangle \\
     & = \delta_{F,F^\prime} \delta_{M,M^\prime}\\
     & \times (-1)^{F_1^\prime+F^\prime+I_H} \sixJ{I_H}{F_1}{F^\prime}{F_1^\prime}{I_H}{1} \sqrt{I_H(I_H+1)(2I_H+1)} \\
     & \times  \sqrt{(2F_1+1)5(2F_1^\prime+1)} \nineJ{N}{N^\prime}{2}{G}{G^\prime}{1}{F_1}{F_1^\prime}{1} \\
     & \times (-1)^{N-K} \sqrt{(2N+1)(2N^\prime+1)} \threeJ{N}{-K}{2}{0}{N^\prime}{K^\prime} \\
     &  \times (-1)^{G + I + 1 + S} \sqrt{(2G+1)(2G^\prime+1)} \sixJ{S}{G^\prime}{I}{G}{S}{1} \sqrt{S(S+1)(2S+1)}
    \end{split}
\end{equation}

The Quadrupole term:

\begin{equation}
    \begin{split}
     \langle K; N, S, I, G, F_1, I_H, F, M |&  T^2(I, I) \cdot T^2(n) | K^\prime; N^\prime, S, I, G^\prime, F_1^\prime, I_H, F^\prime, M^\prime  \rangle \\
    & = \delta_{F,F^\prime} \delta_{M,M^\prime} \delta_{F_1,F_1^\prime} \\
    &  \times (-1)^{N^\prime + F_1 + G} \sixJ{G^\prime}{N^\prime}{F_1}{N}{G}{2}   \\
    & \times (-1)^{N-K} \sqrt{(2N+1)(2N^\prime+1)} \threeJ{N}{-K}{2}{0}{N^\prime}{K^\prime} \\
    &  \times (-1)^{G^\prime + I + 2 + S} \sqrt{(2G+1)(2G^\prime+1)} \sixJ{I}{G^\prime}{S}{G}{I}{2} \\  
    & \times \frac{1}{2 \sqrt{6}} \sqrt{(2I-1)(2I)(2I+1)(2I+2)(2I+3)} 
    \end{split}
\end{equation}

The $\ell$-doubling $q_G$ term:

\begin{equation}
    \begin{split}
     \langle K; N, S, I, G, F_1, I_H, F, M |&  T^2_{2q} (N,N) e^{-2iq\phi} | K^\prime; N^\prime, S, I, G^\prime, F_1^\prime, I_H, F^\prime, M^\prime  \rangle \\
        & = \delta_{F,F^\prime} \delta_{M,M^\prime} \delta_{F_1,F_1^\prime} \delta_{G,G^\prime} \delta_{N, N^\prime}\\
        & \times (-1)^{N-K} \threeJ{N}{-K}{2}{2q}{N^\prime}{K^\prime} \\
        & \times \frac{1}{2 \sqrt{6}} \sqrt{(2N-1)(2N)(2N+1)(2N+2)(2N+3)}\\
    \end{split}
\end{equation}

The $\ell$-doubling $p_G$ term:

Using eq. (2.3.35) in ref.~\cite{hirota_high-resolution_1985}, 

\begin{equation}
    \begin{split} 
     \langle K; N, S, I, G, F_1, I_H, F, M |&   T^2_{2q} (N,S) e^{-2iq\phi}  | K^\prime; N^\prime, S, I, G^\prime, F_1^\prime, I_H, F^\prime, M^\prime  \rangle \\
        & = \delta_{F,F^\prime} \delta_{M,M^\prime} \delta_{F_1,F_1^\prime} \\
        &  \times (-1)^{N^\prime + F_1^\prime + G} \sixJ{G^\prime}{N^\prime}{F_1^\prime}{N}{G}{1}   \\
        &  \times (-1)^{G^\prime + S + 1 + I} \sqrt{(2G+1)(2G^\prime+1)} \sixJ{S}{G^\prime}{I}{G}{S}{1} \sqrt{S(S+1)(2S+1)} \\
        & \times \sqrt{5} \sixJ{1}{1}{2}{N}{N^\prime}{N^\prime} \sqrt{N^\prime(N^\prime+1)(2N^\prime+1)} \\
        & \times (-1)^{N-K} \threeJ{N}{-K}{2}{2q}{N^\prime}{K^\prime} \sqrt{(2N+1)(2N^\prime+1)}\\
    \end{split}
\end{equation}

The Stark term:

\begin{equation}
    \begin{split}
     \langle K; N, S, I, G, F_1, I_H, F, M |&  T^1_{p=0}(D) | K^\prime; N^\prime, S, I, G^\prime, F_1^\prime, I_H, F^\prime, M^\prime  \rangle \\
     & = (-1)^{F-M} \threeJ{F}{-M}{1}{0}{F^\prime}{M^\prime} \\
     & \times (-1)^{F^\prime+F_1+1+I_H} \sqrt{(2F+1)(2F^\prime+1)} \sixJ{F_1^\prime}{F^\prime}{I_H}{F}{F_1}{1}\\
     &\times \delta_{G,G^\prime} (-1)^{F_1^\prime+N+1+G} \sqrt{(2F_1+1)(2F_1^\prime+1)} \sixJ{N^\prime}{F_1^\prime}{G}{F_1}{N}{1}\\
     &\times (-1)^{N-K} \sqrt{(2N+1)(2N^\prime+1)} \threeJ{N}{-K}{1}{0}{N^\prime}{K^\prime}
    \end{split}
\end{equation}

Note that we only consider the contribution from the dipole component along the molecular $z$ axis.  \\ \\

The Zeeman term:

\begin{equation}
    \begin{split}
     \langle K; N, S, I, G, F_1, I_H, F, M |&  T^1_{p=0}(S) | K^\prime; N^\prime, S, I, G^\prime, F_1^\prime, I_H, F^\prime, M^\prime  \rangle \\
    & = (-1)^{F-M} \threeJ{F}{-M}{1}{0}{F^\prime}{M^\prime} \\ 
    & \times (-1)^{F^\prime+F_1+1+I_H} \sqrt{(2F+1)(2F^\prime+1)} \sixJ{F_1^\prime}{F^\prime}{I_H}{F}{F_1}{1}\\ 
    &\times  \delta_{N, N^\prime} \delta_{K,K^\prime} (-1)^{F_1+N+1+G^\prime} \sqrt{(2F_1+1)(2F_1^\prime+1)} \sixJ{G^\prime}{F_1^\prime}{N}{F_1}{G}{1}\\
    &  \times (-1)^{G + I + 1 + S} \sqrt{(2G+1)(2G^\prime+1)} \sixJ{S}{G^\prime}{I}{G}{S}{1} \sqrt{S(S+1)(2S+1)}
    \end{split}
\end{equation}

The eEDM term:

\begin{equation}
    \begin{split}
     \langle K; N, S, I, G, F_1, I_H, F, M |&  T^1(S) \cdot T^1(n) | K^\prime; N^\prime, S, I, G^\prime, F_1^\prime, I_H, F^\prime, M^\prime  \rangle \\
    & = \delta_{F,F^\prime} \delta_{M,M^\prime} \delta_{F_1,F_1^\prime} \\
    &  \times (-1)^{N^\prime + F_1 + G} \sixJ{G^\prime}{N^\prime}{F_1}{N}{G}{1}   \\
    & \times (-1)^{N-K} \sqrt{(2N+1)(2N^\prime+1)} \threeJ{N}{-K}{1}{0}{N^\prime}{K^\prime} \\
    &  \times (-1)^{G^\prime + S + 1 + I} \sqrt{(2G+1)(2G^\prime+1)} \sixJ{S}{G^\prime}{I}{G}{S}{1} \sqrt{S(S+1)(2S+1)}
    \end{split}
\end{equation}

The NSM term:

\begin{equation}
    \begin{split}
     \langle K; N, S, I, G, F_1, I_H, F, M |&  T^1(I) \cdot T^1(n) | K^\prime; N^\prime, S, I, G^\prime, F_1^\prime, I_H, F^\prime, M^\prime  \rangle \\
    & = \delta_{F,F^\prime} \delta_{M,M^\prime} \delta_{F_1,F_1^\prime} \\
    &  \times (-1)^{N^\prime + F_1 + G} \sixJ{G^\prime}{N^\prime}{F_1}{N}{G}{1}   \\
    & \times (-1)^{N-K} \sqrt{(2N+1)(2N^\prime+1)} \threeJ{N}{-K}{1}{0}{N^\prime}{K^\prime} \\
    &  \times (-1)^{G+ S + 1 + I} \sqrt{(2G+1)(2G^\prime+1)} \sixJ{I}{G^\prime}{S}{G}{I}{1} \sqrt{I(I+1)(2I+1)}
    \end{split}
\end{equation}

The NMQM term:

NMQM Hamiltonian in spherical notation is given in ref.~\cite{Ho_YbF_MQM_2023}.

\begin{equation}
    \begin{split}
     \langle K; N, S, I, G, F_1, I_H, F, M |&  T^1(S, T) \cdot T^1(n) | K^\prime; N^\prime, S, I, G^\prime, F_1^\prime, I_H, F^\prime, M^\prime  \rangle \\
    & = \delta_{F,F^\prime} \delta_{M,M^\prime} \delta_{F_1,F_1^\prime} \\
    &  \times (-1)^{G^\prime + F_1 + N} \sixJ{N^\prime}{G^\prime}{F_1}{G}{N}{1}   \\
    & \times (-1)^{N-K} \sqrt{(2N+1)(2N^\prime+1)} \threeJ{N}{-K}{1}{0}{N^\prime}{K^\prime} \\
    & \times \sqrt{(2G+1)3(2G^\prime+1)} \nineJ{S}{I}{G^\prime}{1}{2}{1}{S}{I}{G} \\
    & \times \frac{1}{\sqrt{6}} \sqrt{(2I-1)(2I)(2I+1)(2I+2)(2I+3)} \sqrt{S(S+1)(2S+1)} \\    
    \end{split}
\end{equation}

Note that this matrix element is derived in eq. (14) in ref.~\cite{Ho_YbF_MQM_2023}. Although the final form is slightly different, if the analytical form of $\sixJ{F_1}{F}{I_2}{F}{F_1}{0} $ in eq. (14) is substituted, then it coincides with our result above.

\begingroup
\begin{table*}
\renewcommand{\arraystretch}{0.85}
\centering
\begin{threeparttable}
\caption{\label{tab:lines}
Ground states quantum numbers ($N", G", F_1", \mathcal{P}^\prime$), excited states quantum numbers ($J', F_1', \mathcal{P}^\prime$), observed positions, and residuals of $\tilde{A}^2\Pi(000)-\tilde{X}^2\Sigma^+(01^10)$ band of $^{173}$YbOH. There are in total 61 lines assigned. The fit residual is 9.0 MHz.}
\begin{ruledtabular}
\begin{tabularx}{\textwidth}{lllld}

$N^{''}, G^{''}, F_1^{''}, \mathcal{P}$	&	$J', F_1^{'}, \mathcal{P}$	&	Obs. (cm$^{-1}$)	&	Obs. -- Calc. (MHz)	\\ \hline

1, 2, 1, +   &   1.5, 2, --   &   17004.7905   &   --10.0	\\ 
1, 2, 2, +   &   1.5, 2, --   &   17004.7757   &   10.8	\\ 
1, 2, 2, +   &   1.5, 1, --   &   17004.7823   &   5.4	\\ 
1, 2, 3, +   &   1.5, 3, --   &   17004.7831   &   --5.1	\\ 
1, 2, 3, +   &   1.5, 2, --   &   17004.7841   &   --6.8	\\ 
1, 2, 3, +   &   1.5, 4, --   &   17004.7993   &   --9.3	\\ 
1, 2, 1, --   &   2.5, 2, +   &   17006.2784   &   27.4	\\ 
1, 2, 2, --   &   2.5, 3, +   &   17006.2589   &   --4.8	\\ 
1, 2, 2, --   &   2.5, 2, +   &   17006.2618   &   --8.7	\\ 
1, 2, 2, --   &   0.5, 2, +   &   17003.7993   &   --2.0	\\ 
1, 2, 2, --   &   0.5, 3, +   &   17003.8045   &   --7.2	\\ 
1, 2, 3, --   &   2.5, 3, +   &   17006.2684   &   3.3	\\ 
1, 2, 3, --   &   2.5, 4, +   &   17006.2719   &   2.6	\\ 
1, 2, 3, --   &   0.5, 2, +   &   17003.8090   &   15.7	\\ 
1, 2, 3, --   &   0.5, 3, +   &   17003.8142   &   7.1	\\ 
1, 3, 2, +   &   1.5, 3, --   &   17004.9761   &   --8.8	\\ 
1, 3, 2, +   &   1.5, 1, --   &   17004.9842   &   --3.5	\\ 
1, 3, 2, +   &   1.5, 2, --   &   17004.9772   &   --9.4	\\ 
1, 3, 3, +   &   1.5, 4, --   &   17004.9772   &   11.9	\\ 
1, 3, 4, +   &   2.5, 5, --   &   17005.1454   &   10.5	\\ 
1, 3, 4, +   &   1.5, 3, --   &   17004.9736   &   --1.4	\\ 
1, 3, 4, +   &   1.5, 4, --   &   17004.9898   &   --5.8	\\ 
1, 3, 2, --   &   2.5, 3, +   &   17006.4619   &   8.6	\\ 
1, 3, 2, --   &   0.5, 3, +   &   17004.0075   &   8.3	\\ 
1, 3, 3, --   &   2.5, 3, +   &   17006.4452   &   --7.0	\\ 
1, 3, 3, --   &   2.5, 4, +   &   17006.4487   &   --6.3	\\ 
1, 3, 3, --   &   0.5, 2, +   &   17003.9861   &   13.4	\\ 
1, 3, 3, --   &   0.5, 3, +   &   17003.9910   &   --2.8	\\ 
1, 3, 4, --   &   2.5, 4, +   &   17006.4619   &   1.3	\\ 
1, 3, 4, --   &   2.5, 5, +   &   17006.4786   &   --6.3	\\ 
1, 3, 4, --   &   2.5, 5, +   &   17006.4786   &   --6.9	\\ 
2, 2, 4, +   &   0.5, 3, --   &   17002.3808   &   17.5	\\ 
2, 2, 2, --   &   2.5, 1, +   &   17005.2973   &   0.5	\\ 
2, 2, 2, --   &   2.5, 2, +   &   17005.2923   &   --6.1	\\ 
2, 2, 3, --   &   2.5, 3, +   &   17005.2923   &   12.9	\\ 
2, 2, 3, --   &   2.5, 2, +   &   17005.2951   &   5.4	\\ 
2, 2, 4, --   &   2.5, 4, +   &   17005.2845   &   --7.0	\\ 
2, 2, 4, --   &   2.5, 5, +   &   17005.3013   &   --13.6	\\ 
2, 3, 2, +   &   0.5, 2, --   &   17002.5933   &   --14.2	\\ 
2, 3, 3, +   &   3.5, 4, --   &   17007.4713   &   --10.3	\\ 
2, 3, 4, +   &   3.5, 5, --   &   17007.4800   &   --6.0	\\ 
2, 3, 5, +   &   3.5, 5, --   &   17007.4713   &   --1.6	\\ 
2, 3, 5, +   &   3.5, 6, --   &   17007.4883   &   4.8	\\ 
2, 3, 1, --   &   2.5, 1, +   &   17005.4683   &   1.7	\\ 
2, 3, 1, --   &   2.5, 2, +   &   17005.4634   &   --1.7	\\ 
2, 3, 1, --   &   2.5, 0, +   &   17005.4719   &   14.6	\\ 
2, 3, 2, --   &   2.5, 3, +   &   17005.4689   &   3.7	\\ 
2, 3, 2, --   &   2.5, 2, +   &   17005.4719   &   1.8	\\ 
2, 3, 3, --   &   2.5, 4, +   &   17005.4814   &   5.7	\\ 
2, 3, 4, --   &   2.5, 5, +   &   17005.5018   &   3.4	\\ 
2, 3, 4, --   &   2.5, 5, +   &   17005.5018   &   4.1	\\ 
2, 3, 4, --   &   2.5, 4, +   &   17005.4849   &   6.8	\\ 
2, 3, 5, --   &   2.5, 5, +   &   17005.4930   &   --4.3	\\ 
2, 3, 5, --   &   2.5, 5, +   &   17005.4930   &   --2.9	\\ 
3, 2, 2, --   &   4.5, 3, +   &   17008.3141   &   9.7	\\ 
3, 2, 3, --   &   4.5, 4, +   &   17008.3179   &   --7.3	\\ 
3, 2, 4, --   &   4.5, 5, +   &   17008.3179   &   --22.8	\\ 
3, 3, 4, +   &   3.5, 5, --   &   17006.0102   &   5.7	\\ 
3, 3, 5, +   &   3.5, 6, --   &   17006.0278   &   --1.8	\\ 
    
\end{tabularx}
\end{ruledtabular}
\end{threeparttable}
\end{table*}
\endgroup

\begingroup
\begin{table*}
\centering
\begin{threeparttable}
\caption{\label{tab:lines}
Ground states quantum numbers ($N", G", F_1", \mathcal{P}^\prime$), excited states quantum numbers ($J', F_1', \mathcal{P}^\prime$), observed positions, and residuals of $\tilde{A}^2\Pi(000)-\tilde{X}^2\Sigma^+(01^10)$ band of $^{171}$YbOH. There are in total 37 lines assigned. The fit residual is 6.9 MHz.}
\begin{ruledtabular}
\begin{tabularx}{\textwidth}{lllld}

$N^{''}, G^{''}, F_1^{''}, \mathcal{P}$	&	$J', F_1^{'}, \mathcal{P}$	&	Obs. (cm$^{-1}$)	&	Obs. -- Calc. (MHz)	\\ \hline

1, 0, 1, +   &   1.5, 2, --   &   17005.0986   &   10.2	\\ 
1, 0, 1, +   &   1.5, 1, --   &   17005.1126   &   --3.9	\\ 
1, 0, 1, --   &   2.5, 2, +   &   17006.5996   &   --6.1	\\ 
1, 1, 0, +   &   1.5, 1, --   &   17004.8826   &   --12.7	\\ 
1, 1, 1, +   &   1.5, 2, --   &   17004.8721   &   14.5	\\ 
1, 1, 2, +   &   1.5, 2, --   &   17004.8721   &   --2.4	\\ 
1, 1, 2, +   &   1.5, 1, --   &   17004.8863   &   --11.5	\\ 
1, 1, 0, --   &   0.5, 1, +   &   17003.8918   &   9.5	\\ 
1, 1, 1, --   &   0.5, 1, +   &   17003.8945   &   3.6	\\ 
1, 1, 1, --   &   2.5, 2, +   &   17006.3733   &   --1.6	\\ 
1, 1, 1, --   &   1.5, 2, +   &   17004.0096   &   1.9	\\ 
1, 1, 1, --   &   0.5, 0, +   &   17003.9077   &   --9.1	\\ 
1, 1, 2, --   &   0.5, 1, +   &   17003.8945   &   --2.5	\\ 
1, 1, 2, --   &   2.5, 3, +   &   17006.3588   &   4.7	\\ 
1, 1, 2, --   &   2.5, 2, +   &   17006.3737   &   5.9	\\ 
1, 1, 2, --   &   1.5, 2, +   &   17004.0096   &   --4.2	\\ 
2, 0, 2, +   &   3.5, 3, --   &   17007.6141   &   --0.2	\\ 
2, 0, 2, --   &   2.5, 2, +   &   17005.6182   &   --14.1	\\ 
2, 1, 2, +   &   3.5, 3, --   &   17007.3854   &   --3.4	\\ 
2, 1, 2, +   &   1.5, 2, --   &   17003.8918   &   --3.3	\\ 
2, 1, 3, +   &   3.5, 4, --   &   17007.3750   &   7.4	\\ 
2, 1, 1, --   &   2.5, 2, +   &   17005.3888   &   --0.3	\\ 
2, 1, 2, --   &   2.5, 3, +   &   17005.3752   &   8.9	\\ 
2, 1, 2, --   &   2.5, 2, +   &   17005.3897   &   --4.4	\\ 
2, 1, 3, --   &   3.5, 4, +   &   17005.6326   &   --11.2	\\ 
2, 1, 3, --   &   2.5, 3, +   &   17005.3802   &   8.2	\\ 
2, 1, 3, --   &   2.5, 2, +   &   17005.3947   &   --4.0	\\ 
3, 0, 3, +   &   3.5, 3, --   &   17006.1412   &   --0.9	\\ 
3, 0, 3, --   &   4.5, 4, +   &   17008.6462   &   6.8	\\ 
3, 1, 3, +   &   3.5, 4, --   &   17005.8967   &   2.4	\\ 
3, 1, 4, +   &   3.5, 4, --   &   17005.9044   &   9.2	\\ 
3, 1, 2, --   &   2.5, 2, +   &   17003.9224   &   --0.8	\\ 
3, 1, 3, --   &   4.5, 4, +   &   17008.4166   &   --2.0	\\ 
3, 1, 3, --   &   2.5, 3, +   &   17003.9093   &   6.3	\\ 
3, 1, 4, --   &   4.5, 5, +   &   17008.4085   &   --8.4	\\ 
4, 1, 3, --   &   4.5, 4, +   &   17006.4469   &   --1.7	\\ 
4, 1, 5, --   &   4.5, 5, +   &   17006.4453   &   --2.8	\\ 
    
\end{tabularx}
\end{ruledtabular}
\end{threeparttable}
\end{table*}
\endgroup

